\documentclass[aps,prb,preprint,groupedaddress,draft,
              showpacs,showkeys,a4paper]{revtex4}
\usepackage{amsmath}
\usepackage[final]{graphicx} 
\begin{document}
\title{Antivortices due to competing orbital and paramagnetic
  pair-breaking effects}
\author{U.~Klein}
\affiliation{Johannes Kepler Universit{\"a}t Linz, Institut f{\"u}r
  Theoretische Physik, A-4040 Linz, Austria}
\begin{abstract}
Thermodynamically stable vortex-antivortex structures in a
quasi-two-dimensional superconductor in a tilted magnetic field are
 predicted. For this geometry, both orbital and spin pair-breaking
 effects exist, with their relative strength depending on the tilt
 angle $\theta$.  The spectrum of possible states contains as limits the
 ordinary vortex state (for large $\theta$) and the
 Fulde-Ferrell-Larkin-Ovchinnikov state (for $\theta=0$).  The
 quasiclassical equations are solved near $H_{c2}$ for arbitrary $\theta$
 and it is shown that stable states with coexisting vortices and
 antivortices exist in a small interval close to $\theta=0$. The results
 are compared with recent predictions of antivortices in mesoscopic
 samples.
\end{abstract}
\maketitle
\section{Introduction}
In this paper we shall establish a link between two current topics in
superconductivity which seem unrelated at first sight. These two
topics are antivortices in mesoscopic samples and paramagnetic pair
breaking effects in layered superconductors respectively.

Antivortices are topological singularities of the order parameter with
vorticity opposite to ordinary vortices.  Generally, the phase $\chi$ of
the superconducting order parameter $\psi =|\psi| \exp \imath \chi$ must
change by $2\pi n$, when surrounding an isolated zero of $\psi$ which is
a topologically singular point. The arbitrary integer $n$ is referred
to as topological charge or vorticity. An ordinary vortex in a type II
superconductor has positive vorticity, $n=1$, corresponding to a phase
change by $2\pi$, a total flux of a single flux-quantum $\Phi_0$, and a
diamagnetic behavior of the (screening) currents around the vortex
center. An antivortex, on the other hand, has negative vorticity (in
the simplest case $n=-1$) and carries a negative flux (e.g. $-\Phi_0$).
The currents surrounding the antivortex center have opposite direction
as compared to the ordinary vortex, i. e. they are \emph{paramagnetic}
in nature. The diamagnetic nature of the screening currents in the
ordinary, orbitally-limited, superconducting state is a consequence of
the dominating orbital pair-breaking mechanism.

Notwithstanding this fundamentally unfavorable conditions for the
occurrence of antivortices, several specific configurations designed
to create stable vortex-antivortex structures in orbitally-limited
superconductors, either in thermodynamic
equilibrium~\cite{CCBM,EKLP,MILPEE,MFDM2PRL} or as metastable
states~\cite{GHSHSH}, have been proposed in recent years.  Of
particular interest are theoretical predictions in mesoscopic
superconducting samples ~\cite{CCBM,MFDM2PRL}. Free energy
calculations in the framework of GL-theory, following the original
prediction~\cite{CCBM} of stable vortex-antivertex structures,
revealed~\cite{BONKAB,BAELPEET} that such structures are in fact
\emph{unstable} in type II superconductors; this result was obtained
using a GL-parameter $\kappa$ considerably larger than the critical value
$\kappa=1/\sqrt{2}$ separating type II from type I superconductivity. This
instability is a consequence of the attraction between vortices and
antivortices which consequently annihilate each other in the state of
lowest free energy. It may be qualitatively understood by considering
the interaction between well separated vortices in \emph{bulk}
samples~\cite{KRAMERGL}. In this case, the total interaction between
vortices in GL theory is the sum of two contributions, a repulsive
electrodynamic part and an attractive condensation energy part. The
electrodynamic part exceeds the attractive part for $\kappa > 1/\sqrt{2}$.
But for a vortex-antivortex pair the electrodynamic interaction
changes sign and becomes attractive while the condensation energy
interaction does not change sign and remains attractive. Thus, the
total force between vortices and antivortices becomes attractive.

Nevertheless, the search for stable vortex-antivortex structures in
mesoscopic samples was finally successful~\cite{MFDM2PRL}. A state
with total flux of $2\Phi_0$, carried by a antivortex and three ordinary
vortices in a mesoscopic sample of triangular shape, was found to be
stable in a small region below $T_c$ and below $\kappa = 1/\sqrt{2}$ (In
this part of the $\kappa,T-$ plane, GL-theory still
applies~\cite{KLEIN1}). This effect, which has not yet been verified
experimentally, is essentially a surface effect and cannot be
understood in terms of forces between isolated vortices. Contrary to
the interpretation of the authors~\cite{MFDM2PRL}, the decrease of
$\kappa$ below the critical value $1/ \sqrt{2}$ does \emph{not} lead to a
change in sign of the vortex-antivortex interaction. The latter
remains attractive for arbitrary $\kappa$ since only the electrodynamic
part of the interaction changes sign. Nevertheless, the decrease of
$\kappa$ plays an essential role because it weakens the relative
importance of the vortex interactions in comparison to that of the
confining boundary conditions. The geometric shape (triangular) of the
sample is important since no stable vortex-antivortex structure has
been found~\cite{MERKAB} in thin films of \emph{quadratic} shape.
\section{Paramagnetic pair-breaking as a stabilizing factor for antivortices}
Considering the physical mechanism governing the interaction between
vortices it seems impossible to create periodic structures with
antivortices for values of $\kappa$ much larger than $1/ \sqrt{2}$, deep
in the type II region.  This is in fact true for the purely
orbitally-limited superconducting state. However, as will be shown in
this paper, an exotic type of superconductivity exists, characterized
by the presence of two different competing pair-breaking effects,
where vortex-antivortex structures appear quite naturally for a large
range of GL parameters $\kappa$, not as a surface effect but in an
extended, two-dimensional (2D) periodic "vortex"-lattice.

Cooper pairs may either be broken by the orbital effect or by means of
the interaction between the external field and the magnetic moment due
to the spins of the electrons~\cite{CLOGSTON,CHANDRA}. This spin
pair-breaking effect leads to a \emph{paramagnetic} response of the
superconducting condensate. It is usually much smaller than the
orbital effect but becomes important if the orbital effect can be
suppressed. Clearly, decreasing the magnitude of the orbital effect
favors the stability of antivortices.

Until recently, in most works only the limit of complete suppression
of the orbital effect was considered. In this paramagnetic limit, a
spatially inhomogeneous superconducting state, referred to as FFLO
state, has been predicted theoretically by Fulde and
Ferrell~\cite{FUFE}, and by Larkin and Ovchinnikov~\cite{LAOV}. With
regard to the present problem, the FFLO state itself is not a good
candidate for observing antivortices since the diamagnetic response is
\emph{completely} suppressed, and no (anti)vortices at all can exist
in the purely paramagnetic limit.

Obviously, coexistence of \emph{both} pair-breaking mechanisms, with
only partial suppression of the diamagnetic effect, provides optimal
conditions for the stability of vortex-antivortex structures. Such a
situation may be realized by slightly tilting the applied magnetic
field, by an angle $\theta$, from the plane-parallel direction .  If the
conducting layers are completely decoupled and infinitely thin (it has
been shown recently that single-atomic layers are required for
negligible orbital pair-breaking~\cite{MANKL2}) one has, in fact, a
mixture of both pair-breaking mechanisms, with the orbital effect
entirely due to the perpendicular component $H_{\perp}=H\sin\theta$ and
the spin effect mainly due to the parallel component $H_\|$.  The
relative weight of the two pair-breaking effects may be controlled
with the help of the angle $\theta$.

The second order phase boundary $H_{c2}$ for this situation has been
calculated by Bulaevskii~\cite{BULATILT} for $T=0$ and by Shimahara
and Rainer~\cite{SHIMARAIN} for arbitrary $T$. They found a
non-monotonic upper critical field curve with different pieces of the
curve belonging to different Landau quantum numbers $n$ ($n=0,1,\ldots$).
For large $\theta$ one recovers the usual orbital result $n=0$.  With
decreasing $\theta$ unusual pairing states with nonzero $n=1,2,\ldots$ appear
for $\theta$ smaller than a critical angle of the order of one degree. For
$\theta\to0$, one obtains $n\to\infty$ and the upper critical field of the FFLO
state is recovered.

\section{Outline of calculation}

The states below the non-monotonic upper critical field curve which
belong to the lowest nonzero quantum numbers, say $n=1-4$, are most
interesting from the present point of view of antivortices.

We shall make only two very general (and reasonable) assumptions with
regard to the structure of these states. The first is, that the order
parameter modulus $|\Delta|$ and the local magnetic induction $\vec{B}$
are periodic with respect to an arbitrary two-dimensional lattice. The
second assumption is that a unit cell of this quasi-periodic lattice
carries a single flux quantum $\Phi_0$. To find the equilibrium
structure of these states, all possible lattices must be considered
and the global minimum of the free energy must be identified.

Considering the vicinity of the upper critical field, where a second
order transition occurs, a free energy expansion, up to terms of
fourth order, may be performed. Then, the general scheme of such a
calculation is the same as Abrikosov's classical work on type II
superconductivity~\cite{ABRI1}. However, microscopic equations should
be used here, because a GL-formulation, with a finite number of
gradient terms, is not valid in the present (low-temperature)
situation.  The quasiclassical theory of superconductivity,
generalized with regard to Zeeman coupling terms, provides an
appropriate theoretical framework~\cite{KL2DTILT} for the present
problem.

The nonlinear transport equations for a clean superconductor,
including Zeeman coupling, are given by
\begin{equation}
  \label{eq:quasiffp}
\begin{split}
  \left[ 2\bar{\omega}_l+\hbar \vec{v}_F(\vec{k})\vec{\vartheta}_r \right]
  f(\vec{r},\vec{k},\bar{\omega}_l) & = 2\Delta(\vec{r})
  g(\vec{r},\vec{k},\bar{\omega}_l)\mbox{,} \\
  \left[ 2\bar{\omega}_l - \hbar \vec{v}_F(\vec{k}) \vec{\vartheta}_r^{\ast}\right]
  f^{\it+}(\vec{r},\vec{k},\bar{\omega}_l) & = 2\Delta^\ast(\vec{r})
  g(\vec{r},\vec{k},\bar{\omega}_l) \mbox{,}
\end{split}     
\end{equation}
where $f,\,f^+,\,g$ are the quasiclassical Green`s functions
($g^2=1-ff^+$), $\Delta$ is the order parameter, and $\vec{\vartheta}_r$ is the
gauge-invariant derivative, defined by
$\vec{\vartheta}_r=\vec{\nabla}_r-\imath(2e/\hbar c) \vec{A}$. The latter term,
containing the vector potential $\vec{A}$, describes the familiar
orbital coupling between Cooper pairs and magnetic field. In addition
the coupling between the spins of the electron and the magnetic field
is taken into account in Eq.~(\ref{eq:quasiffp}). The corresponding
Zeeman term $\mu B$ is contained in the complex variable
$\bar{\omega}_l=\omega_l+\imath\mu B$ which replaces the real Matsubara
frequency $\omega_l$. The quantity $\vec{v}_F(\vec{k})$ denotes the Fermi
velocity which depends on the quasiparticle wave-number $\vec{k}$.

The order parameter $\Delta$ is defined in terms of the Green`s functions
$f,\,f^+$ by the relation
\begin{equation}
  \label{eq:scop}
\left(2 \pi k_B T  \sum_{l=0}^{N_D} \frac{1}{\bar{\omega}_l}+ 
\ln\left(T/T_c\right) \right) 
\Delta(\vec{r}) = 
\pi k_B T  \sum_{l=0}^{N_D} 
\oint d^2k' \, \left[ f(\vec{r},\vec{k}^\prime,\bar{\omega}_l) +
f(\vec{r},\vec{k}^\prime,\bar{\omega}_l^\ast ) \right] 
\mbox{,}
\end{equation}
where $N_D$ is the cutoff index for the Matsubara sums.  To calculate
the vector potential $\vec{A}$ for given Green`s function $g$,
Maxwell`s equation
\begin{equation}
  \label{eq:scvecpot}
\vec{\nabla}_r \times \left( 
\vec{B}(\vec{r}) + 4\pi \vec{M}(\vec{r}) \right)= 
\frac{16\pi^2ek_BTN_F}{c} \sum_{l=0}^{N_D} 
\oint \frac{d^2k'}{4\pi} \, \vec{v}_F(\vec{k}^\prime)
\Im g(\vec{r},\vec{k}^\prime,\bar{\omega}_l ) \mbox{,} 
\end{equation}
has to be solved. Here, $N_F$ is the normal-state density of states at
the Fermi level. The r.h.s. of Eq.~(\ref{eq:scvecpot}) is the familiar
(orbital) London screening current while the magnetization $\vec{M}$
is a consequence of the magnetic moments of the electrons and is given
by
\begin{equation}
  \label{eq:magn}
\vec{M}(\vec{r})=
2\mu^2N_F\vec{B}(\vec{r}) 
-4\pi k_BTN_F \mu \sum_{l=0}^{N_D} \oint  \frac{d^2k'}{4\pi} 
\Im g \frac{\vec{B}}{B} 
\mbox{,} 
\end{equation}  
The first term on the r.h.s. of Eq.~(\ref{eq:magn}) is the normal
state spin polarization. The second term is is a spin polarization due
to quasiparticles in the superconducting state.

Solving these equations one obtains extrema of a free energy
functional $G$ (not written down here, see~\cite{KL2DTILT}), whose
global minima represent the stable states. The quasiclassical
equations ~(\ref{eq:quasiffp})-(\ref{eq:magn}) are rather general;
they describe both orbital and paramagnetic pair-breaking effects, and
cover the ordinary vortex state as well as the FFLO state in the
appropriate limits. In addition,
Eqs.~(\ref{eq:quasiffp})-(\ref{eq:magn}) take into account the
magnetic response of the superconductor (arbitrary values of the
GL-parameter $\kappa$ may be considered), which is neglected in most
treatments of paramagnetic pair breaking. Note that the familiar
London screening current is completely absent in the purely
paramagnetic (FFLO) limit.

The calculation generalizes techniques used previously to calculate
the upper critical field~\cite{BULATILT},~\cite{SHIMARAIN} and the
equilibrium structure in the high-$\kappa$-limit~\cite{KLRAISHI}. For the
vector potential the same gauge as in previous numerical
calculations~\cite{KLEIN1,KLEIN2} on the vortex lattice (without
Zeeman coupling) may be used. A first important step is the solution
of the transport equation~(\ref{eq:quasiffp}) for small $|\Delta|$, taking
derivatives of $|\Delta|$ of arbitrary order into account . This is
achieved by means of an eigenfunction expansion and the so-called
``Helfand-Werthammer Trick'', leading to an integral representation
for the first-order Greens functions. In the course of the following
free energy expansion, up to terms of fourth order in $|\Delta|$, a large
number of momentum- and configuration space integrations have to be
performed. The evaluation of these integrals may be greatly simplified
by introducing the gap correlation function. A more detailed
description of the lengthy calculation (and a discussion of other
results, not related to antivortices) may be found
elsewhere~\cite{KL2DTILT}.

\section{Stable vortex-antivortex structures for n=2}
In this section results are reported which show that stable
vortex-antivortex structures do in fact exist in bulk superconductors
for nonzero finite $n$. The most important of these states, which is
most stable and most easily accessible experimentally, belongs to
Landau quantum number $n=2$. Discussion will be restricted here to
this single state (Stable structures for other $n$ and a discussion of
the general properties of the new paramagnetic vortex states have been
reported elsewhere~\cite{KL2DTILT}).
\begin{figure}[htbp]  
\begin{center}\leavevmode
\includegraphics[width=8cm]{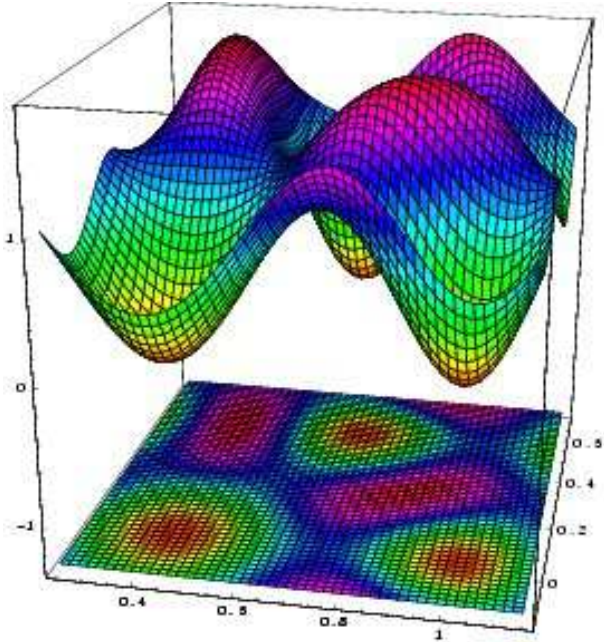}
\caption{\label{fig:opn2}   Square of 
modulus of order parameter $|\psi_2|^2$ for Landau level $n=2$ 
as a function of $x/a,\,y/a$ in the range 
$0.25<x/a<1.25,\;-0.15<y/a<0.7$. This is the stable 
structure (unit cell parameters $a/b=1,\,\alpha=60\mbox{{\r{}}}$) for 
$\mu=0.1,\,\tilde{\kappa}=10,\,\theta=0.7\mbox{{\r{}}},\,t=0.2$.}
\end{center} 
\end{figure}
The results depend on $4$ input parameters, the dimensionless magnetic
moment $\mu=\pi k_BT_c/mv_F^2$, Eilenberger's GL-parameter $\tilde{\kappa}$
(which is defined in terms of the GL-parameter $\kappa$ of a clean
superconductor by the relation $\tilde{\kappa}=0.6837\kappa$), the tilt angle
$\theta$, and the dimensionless temperature $t=T/T_c$. Given these
numbers, the critical field $B_{c2}$, Landau's quantum number $n$, the
unit-cell structure of the stable lattice (which is characterized by
the lengths $a,\,b $ of the unit-cell vectors and by the angle $\alpha$
between them), and the fields $\psi_{n}(\vec{r}),\,\vec{B}(\vec{r})$ may
be calculated~\cite{KL2DTILT}. The square of the order parameter
modulus for pairing in Landau level $n$ is given by the Fourier series
\begin{equation}
  \label{eq:opmodsqare}
|\psi_{n}|^2(\vec{r})= \sum_{l,j} (\psi_{n}^2)_{l,j}\textrm{e}^{\imath \vec{Q}_{l,j}\vec{r}}
\mbox{,}
\end{equation}  
with reciprocal lattice vectors $\vec{Q}_{l,j}$ corresponding to the
lattice parameters $a,\,b,\,\alpha$, and with Fourier coefficients
\begin{equation}
  \label{eq:opmsfc}
(\psi_{n}^2)_{l,j}=(-1)^{lj} \textrm{e}^{-\imath \pi l \frac{b}{a}\cos\alpha } 
\mathrm{e}^{-x_{l,j}/2} L_n(x_{l,j})
\mbox{.}
\end{equation}
The quantity $x_{l,j}$ is defined elsewhere~\cite{KL2DTILT}.  The
magnetic response of the superconductor is characterized by means of
the deviation $\vec{B}_1=\vec{B}-\bar{\vec{B}}$ of the local induction
from the macroscopic induction $\bar{\vec{B}}$. The parallel component
of $\vec{B}_1$, which describes a purely paramagnetic behavior, is
denoted by $B_{1\parallel}$. The perpendicular component $B_{1\perp}$
describes a mixed orbital and paramagnetic response.  The Fourier
coefficients of these quantities are given by
\begin{align}
  (B_{1\parallel})_{l,m}=&-\frac{t\langle|\Delta_n|^2\rangle}{\tilde{\kappa}^2-\mu^2}
  \bar{B}_{\parallel} \mu^2 \cdot  \nonumber \\
  &(-1)^{lm} \textrm{e}^{-\imath \pi l \frac{b}{a}\cos\alpha } f_1(x_{l,m})
  S_{l,m}^{(1)} \mbox{,}
\label{eq:b1par}  \\
(B_{1\perp})_{l,m}=& -\frac{t\langle|\Delta_n|^2\rangle}{\tilde{\kappa}^2-\mu^2}
(-1)^{lm} \textrm{e}^{-\imath \pi l \frac{b}{a}\cos\alpha }  \cdot  \nonumber \\
&\left(\bar{B}_{\perp } \mu^2 f_1(x_{l,m}) S_{l,m}^{(1)} - g_1(x_{l,m})
  S_{l,m}^{(2)}\right)
\label{eq:b1senk}  
\mbox{,}
\end{align}
where $\langle|\Delta_n|^2\rangle$ denotes the spatial average of the square of the
order parameter and the functions $f_1,\,g_1$ and the Matsubara sums
$S_{l,m}^{(i)}$ are reported elsewhere~\cite{KL2DTILT}.
\begin{figure}[htbp] 
\begin{center}\leavevmode
\includegraphics[width=8cm]{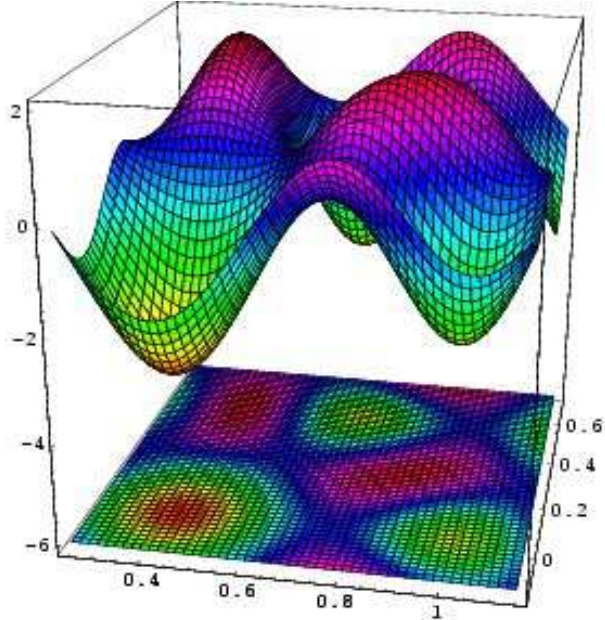}
\caption{\label{fig:bpar1n2}   Spatially 
varying part of the parallel magnetic field 
$B_{1\parallel}(\vec{r})$ using the same input parameters and 
the same part of the $x,\,y$-plane as in 
Figure~\ref{fig:opn2}.}
\end{center} 
\end{figure}
\begin{figure}[htbp] 
\begin{center}\leavevmode
\includegraphics[width=8cm]{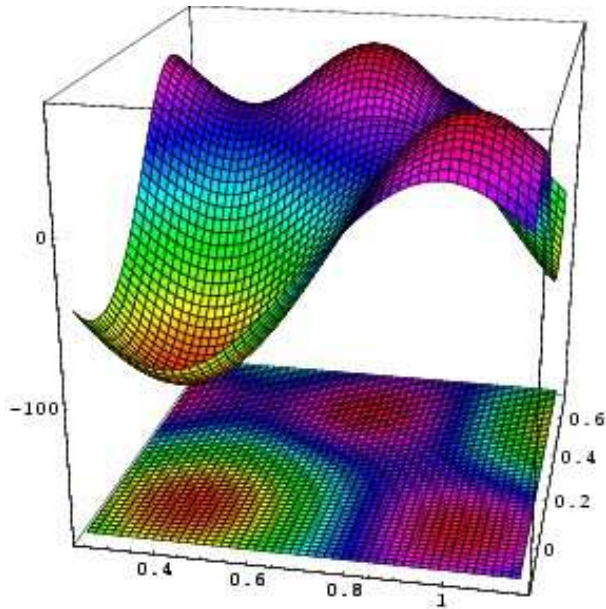}
\caption{\label{fig:bsenk1n2}   Spatially 
varying part of the perpendicular component of the 
magnetic field $B_{1\perp}(\vec{r})$ using the same input 
parameters and the same part of the $x,\,y$-plane as 
in Figure~\ref{fig:opn2}.}
\end{center} 
\end{figure}
Choosing parameters $\mu=0.1$, $\tilde{\kappa}=10$, $\theta=0.7\mbox{{\r{}}}$,
and $t=0.2$ one obtains a Landau quantum number $n=2$
($B_{c2}=4.27086$ if measured in Eilenberger units~\cite{KL2DTILT})
and a unit cell with $a/b=1,\,\alpha=60\mbox{{\r{}}}$, i.e. a triangular
lattice. The order parameter has three zeros per unit cell as shown in
Figure~\ref{fig:opn2}. Thus, one expects that this stable structure
contains two vortices and a single antivortex, because the total flux
per unit-cell is $\Phi_0$.
 
The nature of these order parameter zeros may be further clarified
with the help of the local magnetic field. We plot the fields
$B_{1\parallel}$ and $B_{1\perp}$ as given by
equations.~(\ref{eq:b1par})-(\ref{eq:b1senk}), omitting a common
factor $t\langle|\Delta_n|^2\rangle/(\tilde{\kappa}^2-\mu^2)$ and using the same input
parameters and the same part of the $x,\,y$-plane as in
Figure~\ref{fig:opn2}.  Let us discuss first the parallel component
$B_{1\parallel}$. As shown in figure~\ref{fig:bpar1n2}, the functional
form of $B_{1\parallel}$ is nearly identical with that of $\psi_{2}$
plotted in Figure~\ref{fig:opn2}. This behavior, though unexpected at
first sight, is a consequence of the purely paramagnetic pair-breaking
mechanism for the parallel component. The latter implies an
\emph{enhancement} of the local induction at points of enhanced order
parameter, which is opposite to the familiar diamagnetic field
expulsion.

The spatial dependence of the perpendicular component $B_{1\perp}$,
plotted in Figure~\ref{fig:bsenk1n2}, shows that the order parameter
zero on the left hand side of Figure~\ref{fig:opn2} is an antivortex
while the two remaining zeros belong to ordinary vortices.  At the
centers of the ordinary vortices the field has maxima, i.e. shows the
familiar behavior of orbitally-limited superconductors. At the center
of the antivortex the field has a minimum, i.e. shows the opposite,
paramagnetic behavior. Thus, stable vortex-antivortex structures may
indeed be produced by means of a 2D-superconductor in a tilted
magnetic field. It is the simultaneous presence of both pair-breaking
mechanisms which allows the coexistence of vortices and antivortices
in a periodic lattice.  Calculations~\cite{KL2DTILT} for $\kappa =
0.1,\,1,\,100$ and $t=0.2,\,0.5$ show that the structure displayed in
Figures~\ref{fig:opn2}-\ref{fig:bsenk1n2} is stable for a large range
of parameters.

Comparing the present results with previous predictions, one notes
that the present unit cell has the same shape as the mesoscopic sample
used in the work of Misko et al.~\cite{MFDM2PRL}; the number and type
of order parameter zeros is, however, different. A second common
feature is the strong interaction (small distance) between
anti/vortices.  The basic mechanism responsible for the stability of
antivortices is, however, completely different in both cases.
Experimental verification of the structure predicted here requires
similar conditions as for the FFLO
state~\cite{HIR5,MANKL,NSSBAP,RADOVAN}, namely a layered
superconducting material of high purity with nearly decoupled
conducting planes and a very accurate adjustment of the direction of
the applied magnetic field. Appropriate materials include the
intercalated transition metal dichalcogenide $TaS_2-(pyridine)$, the
organic superconductor $\kappa-(BEDT-TTF)_2Cu(NCS)_2$, or the magnetic
field induced superconductor $\lambda-(BETS)_2FeCl_4$. The predicted
phenomena do not sensitively depend on $\kappa$ and $\mu$ but large values
of these parameters are favorable, because of an associated
enlargement of the paramagnetic region. A non-monotonic phase boundary
in a tilted magnetic field has been observed recently~\cite{RADOVAN}
in the heavy-fermion superconductor $CeCoIn_5$. These data may be
related to the present prediction, despite the fact that $CeCoIn_5$
differs considerably from the simple model considered here.

\begin{acknowledgments}
  I would like to thank V. V. Moshchalkov for a discussion about his
  work~\cite{MFDM2PRL} on antivortices in mesoscopic samples.
\end{acknowledgments}

\bibliography{ffloanti}

\end{document}